\documentclass[twocolumn,twocolappendix]{aastex701}

\usepackage{multirow,amssymb,amsmath,amsfonts}

\newcommand{\rev}[1]{#1}

\newcommand{\rsp}{R_{\mathrm{sp}}} 
\newcommand{\rid}{R_{\mathrm{id}}}
\newcommand{\rvir}{R_{\mathrm{vir}}}
\newcommand{\mvir}{M_{\mathrm{vir}}}
\newcommand{\vvir}{V_{\mathrm{vir}}}
\newcommand{\rtwom}{R_{\mathrm{200m}}}
\newcommand{\Gdyn}{\Gamma_{\mathrm{dyn}}}
\newcommand{\hM}{h^{-1}\mathrm{M}_\odot}
\newcommand{\Om}{\Omega_{\mathrm{m}}}

\begin{document}


\shorttitle{Depletion radius v.s. accretion history}
\shortauthors{Zhou \& Han}

\title{Accretion history dependence of the halo depletion radius}

\author{Jiale Zhou}
\affil{Department of Astronomy, School of Physics and Astronomy, Shanghai Jiao Tong University, Shanghai, 200240, China}
\affil{State Key Laboratory of Dark Matter Physics, Key Laboratory for Particle Astrophysics and Cosmology (MOE), \& Shanghai Key Laboratory for Particle Physics and Cosmology, Shanghai Jiao Tong University, Shanghai 200240, China}
\email{jialezhou@sjtu.edu.cn}

\author[0000-0002-8010-6715]{Jiaxin Han}
\affil{Department of Astronomy, School of Physics and Astronomy, Shanghai Jiao Tong University, Shanghai, 200240, China}
\affil{State Key Laboratory of Dark Matter Physics, Key Laboratory for Particle Astrophysics and Cosmology (MOE), \& Shanghai Key Laboratory for Particle Physics and Cosmology, Shanghai Jiao Tong University, Shanghai 200240, China}
\email{jiaxin.han@sjtu.edu.cn}

\correspondingauthor{Jiaxin Han}
\email{jiaxin.han@sjtu.edu.cn}

\begin{abstract}
We investigate the role of the accretion history in shaping the depletion radius of dark matter halos using a large cosmological N-body simulation. We show that the inner depletion radius, rescaled by the virial radius, depends strongly on the recent mass accretion rate (MAR) measured over a dynamical timescale, while exhibiting only weak dependence on halo mass. While this dependence mirrors that of the splashback radius and the two radii are tightly correlated, the depletion radius exhibits a more nuanced response to the detailed accretion mode. Specifically, we find that the dependence on MAR steepens at lower redshifts, aligning with self-similar spherical collapse models yet contrasting with the behavior of the splashback radius. This redshift dependence is largely driven by recent major mergers, \rev{which tend to suppress the depletion radius in stacked measurements at low redshift and high accretion rate.} Furthermore, we identify a dichotomy in the drivers of the depletion radius. For slowly accreting halos, the MAR is the primary dependence, whereas for rapidly accreting halos, other properties (shape, spin, concentration, and formation time of the central subhalo) related to the anisotropic or perturbed accretion mode also play a significant role. These results establish the depletion radius as a sensitive physical probe of the detailed accretion history of dark matter halos, complementary to the splashback radius.
\end{abstract}

\section{Introduction} \label{sec:intro}

Dark matter halos provide the fundamental building blocks of cosmic structure formation and galaxy evolution. However, a physically motivated definition of their outer boundary remains under active investigation. The difficulty arises from the gradual transition between the virialized region and the surrounding large scale environment, making the characterization of halo boundaries intrinsically ambiguous.

A widely adopted definition is the spherical overdensity radius, motivated by the spherical collapse model \citep{1972ApJ...176....1G}. In this framework, a uniform spherical perturbation collapses under self gravity and reaches virial equilibrium with a characteristic overdensity relative to the cosmic mean matter density. The corresponding boundary, commonly referred to as the virial radius, is routinely used to identify halos in simulations and observations. Despite its practical convenience, this definition has well-known limitations. The overdensity threshold is somewhat arbitrary, the enclosed mass is subject to pseudo evolution \citep{diemer2013pseudo}, and dynamically associated material can extend beyond the nominal virial radius. These issues have motivated ongoing efforts to identify halo boundaries that more directly reflect the dynamical state and growth of halos.

A major development in this direction is the splashback radius ($\rsp$), proposed as a dynamical boundary associated with the first apocenter of recently accreted matter \citep{DK14, More15}. In the self-similar spherical collapse model \citep{1984ApJ...281....1F}, it corresponds to the outermost caustic and is associated with the steepest slope in the density profile. Simulations have shown that $\rsp$ primarily depends on the recent mass accretion rate and redshift, broadly consistent with spherical collapse predictions \citep{Adhikari14, Shi16}.

A complementary dynamical boundary, the depletion radius, was introduced by \citet{FH21}. Rather than tracing particle orbits, the depletion radius is defined through the time evolution of the density profile. As halos accrete matter, material is drawn from their surroundings, producing an outer region where the density declines over time. The depletion radius\footnote{We use the depletion radius to refer to the inner depletion radius defined in \citet{FH21}.} ($\rid$) marks the transition where the local density evolution changes sign, separating the inner growth region from the outer depletion region. This boundary has been identified in simulations \citep{FH21} and detected observationally through weak lensing and satellite kinematics \citep{Fong22, 2021ApJ...915L..18L}. Intriguingly, it has been reported to lie at approximately twice the virial radius across a wide range of halo masses and redshifts \citep{Gao23}.

Despite these advances, several key questions remain open. In the spherical collapse picture, the depletion radius is expected to coincide with the outermost caustic \citep{Han26}, implying a primary dependence on the mass accretion rate and redshift, similar to that of the splashback radius. However, the quantitative dependence of $\rid$ on the recent accretion rate, halo mass, redshift, and secondary halo properties has not yet been systematically characterized.

In this work, we use a large cosmological N-body simulation to systematically investigate the dependence of the depletion radius on halo accretion history and other halo properties, quantify its dependence on the recent accretion rate and redshift, and establish its connection to the splashback radius.

\rev{We note that previous studies have shown that the inferred splashback radius depends on the tracer population, with measurements based on dark matter particles and subhalos corresponding to different percentiles of the particle splashback distribution \citep[e.g.][]{Xhakaj20}. In recent years, hydrodynamical simulations have increasingly been used to investigate dynamical halo boundaries and their observational manifestations using observable tracers such as galaxies, gas, and stars \citep[e.g.,][]{Baxter21, Deason21, ONeil21, Dacunha22, Pizzardo24, OShea25}. Similar tracer dependence may also exist for the depletion radius. However, in this work, we focus on the depletion radius measured from dark matter particles in collisionless simulations, which provides a clean comparison with the theoretical model. The impact of baryonic physics and different tracers on the depletion radius will be explored in future work.}

The paper is organized as follows. Section~\ref{sec:data} describes the simulation and the measurement of halo boundaries. Section~\ref{sec:mar_dependence} presents the dependence of the depletion radius on the mass accretion rate, halo mass, and redshift. Section~\ref{sec:discussions} discusses the role of major mergers and secondary halo properties, and provides an explanation for the approximately constant ratio $\rid/\rvir \simeq 2$. Our conclusions are summarized in Section~\ref{sec:conclusions}.

\section{Data}\label{sec:data}

\subsection{Simulation and the depletion catalog}\label{subsec:data_halocat}
We use a large cosmological N-body simulation drawn from the CosmicGrowth suite \citep{2019SCPMA..6219511J}, which is the same simulation analyzed in \citet{FH21}. The simulation evolves $3072^3$ dark matter particles in a periodic box of side length $600\ h^{-1}\mathrm{Mpc}$, corresponding to a particle mass of $m_{\mathrm{p}} = 5.54 \times 10^8\ \hM$. The adopted cosmology is a flat $\Lambda$CDM model with $\Om=0.268$ and $\Omega_\Lambda=0.732$. Halos are identified using the friends-of-friends (FoF) algorithm \citep{1985ApJ...292..371D} with linking length $b=0.2$, while subhalos and merger trees are constructed with the \textsc{HBT+} code \citep{2012MNRAS.427.2437H,2018MNRAS.474..604H}\footnote{\url{https://github.com/Kambrian/HBTplus}}.

Throughout this paper, we denote by $R_{\Delta \mathrm{x}}$ the radius enclosing a mean density $\Delta\rho_{\mathrm{x}}$, where $\Delta\rho_{\mathrm{x}}$ may correspond to $200\rho_{\mathrm{m}}$, $200\rho_{\mathrm{c}}$, or $\rho_{\mathrm{vir}}$. Here $\rho_{\mathrm{m}}$ and $\rho_{\mathrm{c}}$ are the mean matter and critical densities of the Universe, respectively, and $\rho_{\mathrm{vir}}$ follows the fitting formula for the virialized spherical collapse solution in a flat $\Lambda$CDM cosmology \citep{1998ApJ...495...80B}. Unless otherwise specified, we adopt $\rvir$ as our virial radius, and $\mvir$ as the halo mass. The corresponding virial velocity is given by $\vvir=\sqrt{G\mvir/\rvir}$.

Our final halo sample is constructed using three selection criteria.
(1) We require a minimum virial mass of $\log_{10}[\mvir/(\hM)]\geq12$, corresponding to approximately 1800 particles within $\rvir$, in order to ensure sufficient resolution for reliable profile measurements. 
(2) To maintain a well-defined evolutionary history, we require that the central subhalo of each system remains identified as the central for at least one dynamical time prior to the snapshot under consideration, where the dynamical time is defined as $t_{\mathrm{dyn}}=2\rvir/\vvir$.
(3) To minimize environmental contamination from neighboring systems, \rev{we exclude halos whose exclusion regions overlap with those of more massive halos, where the exclusion radius of each halo is defined as $2.5\rvir$.} This radius exceeds the typical depletion radius ($\simeq 2\rvir$), thereby reducing potential overlap in the outskirts and ensuring that the depletion feature is not significantly affected by nearby massive structures.

We apply these selection criteria independently at 9 snapshots spanning $0 \le z \le 3$ to investigate redshift evolution under a consistent sample definition, \rev{removing approximately 25\% of halos from the original FoF catalog}. The resulting halo catalog contains approximately \rev{$5.6\times10^{5}$} halos at $z=0$, decreasing to $9.7\times10^{4}$ at $z=3$. For subsequent analysis, halos are divided into 5 logarithmic mass bins uniformly spaced between $\log_{10}[\mvir/(\hM)]=12$ and $14.5$. \rev{In the following analysis, the halo sample after applying the above selection criteria is referred to as ``clean sample", to distinguish from the ``smooth sample" we define in the discussion section~\ref{subsec:major_merger} that excludes major merger systems. The number of halos in each bin is listed in Table~\ref{tab:halo_number}.}

\subsection{Halo profiles and boundaries}\label{subsec:data_radius}

The depletion radius can be derived directly from the continuity equation for matter flow. It marks the radius where the local density evolution changes sign, i.e., where $\partial \rho(r)/\partial t=0$. According to the continuity equation in spherical symmetry, the density evolution at radius $r$ is
\begin{equation}
    \frac{\partial \rho(r)}{\partial t}=-\frac{1}{4\pi r^2}\frac{\partial \mathrm{MFR(r)}}{\partial r}, 
\end{equation}
where the mass flow rate is defined as $\mathrm{MFR}(r)=4\pi r^2 \rho(r) v_r(r)$. Thus, the depletion radius corresponds to the location of a local minimum in the MFR profile.

For each halo, we compute the spherically averaged density profile $\rho(r)$, the radial velocity profile $v_r(r)$ (including the Hubble flow), and the MFR profile in 66 logarithmically spaced radial bins from $0.1\rvir$ to $10\rvir$, centering on the most bound particle of the central subhalo. We stack halos with similar properties (e.g., virial mass, mass accretion rate) and extract the median in each radial bin. \rev{Throughout this paper, only bins containing more than 30 halos are shown, in order to ensure statistical robustness and reduce noise.} For consistency across halo masses and redshifts, all profiles are normalized by the corresponding virial quantities of each halo prior to stacking. Specifically, the density profile is expressed in units of the virial density $\rho_{\mathrm{vir}}$, the radial velocity in units of the virial velocity $\vvir$, and the mass flow rate in units of $\mvir \vvir \rvir^{-1}$. Radial distances are scaled by $\rvir$. This normalization removes the leading order self-similar scaling with halo mass and redshift, enabling a direct comparison of structural differences associated with accretion history. 

The logarithmic slope of the density profile, $\mathrm{d} \log \rho(r)/\mathrm{d} \log r$, is obtained by smoothing the stacked median logarithmic density profile with a Savitzky-Golay filter \citep{1964AnaCh..36.1627S} with a polynomial order of 4 and a window length of 15, following \citet{DK14}. The resulting slope profiles typically exhibit a clear splashback feature. We locate the minimum of the curve and fit a quadratic function to the minimum and its two neighboring points to obtain the fitted splashback radius $\rsp$ in units of $\rvir$. The same smoothing and quadratic fitting procedure is applied to the MFR profiles to suppress residual noise and mitigate binning effects. The depletion radius $\rid$ is defined as the location of the fitted minimum of the smoothing MFR profile, also expressed in units of $\rvir$. Moderate variations in the smoothing parameters do not affect our main results.

When applying the Savitzky-Golay filter, the smoothing window requires data points on both sides of each radial bin. As a result, the innermost and outermost 7 radial bins cannot be smoothed for the adopted window length of 15 bins. However, the splashback and depletion radii lie well within the radial range unaffected by these edge effects. We therefore truncate the smoothed profiles accordingly. In this work, all density slope and MFR profiles presented are smoothed and shown only within the valid radial range, unless otherwise specified.

\subsection{Mass accretion rate}\label{subsec:data_mar}
We adopt the same definition of the mass accretion rate as in \citet{DK14}, defined as the logarithmic growth rate of the virial mass with respect to the scale factor, $\Gamma \equiv \Delta \log \mvir / \Delta \log a$. This accretion rate can be computed between different time intervals to probe halo growth on different physical timescales. In this work, we compute $\Gamma$ over one dynamical timescale, selecting the progenitor snapshot closest to one dynamical time prior to the snapshot under consideration. The earlier mass is taken to be the virial mass of the main progenitor halo identified through the merger tree. We denote this accretion rate as $\Gdyn$. This choice captures the recent mass growth relevant for the current dynamical state of the halo.

\rev{We note that previous foundational splashback studies \citep[e.g.,][]{DK14, More15, Diemer17, Mansfield17} commonly adopted $\rtwom$ as the halo radius used for rescaling and defined the accretion rate using different overdensity masses and time intervals. In this work, we instead use $\rvir$ and define $\Gdyn$ over one dynamical time based on $\mvir$. These differences may lead to quantitative offsets when comparing directly with previous splashback results. Nevertheless, previous work has shown that the splashback radius remains strongly correlated with accretion rates defined using reasonable variations of halo mass and time intervals, although with varying correlation strengths \citep{Shin23}. As we will show below, despite these differences, the overall trends remain qualitatively consistent with previous splashback studies.}

\section{Depletion radius dependence on the mass accretion rate}\label{sec:mar_dependence}
According to the spherical collapse model \citep{Adhikari14, Shi16}, the location of the outermost caustic depends only on the mass accretion rate and redshift. We test this prediction using our simulation.

\subsection{Dependence on mass accretion rate}\label{subsec:mar_dependence}

\begin{figure*}[t!]
\centering
\includegraphics[width=1.8\columnwidth]{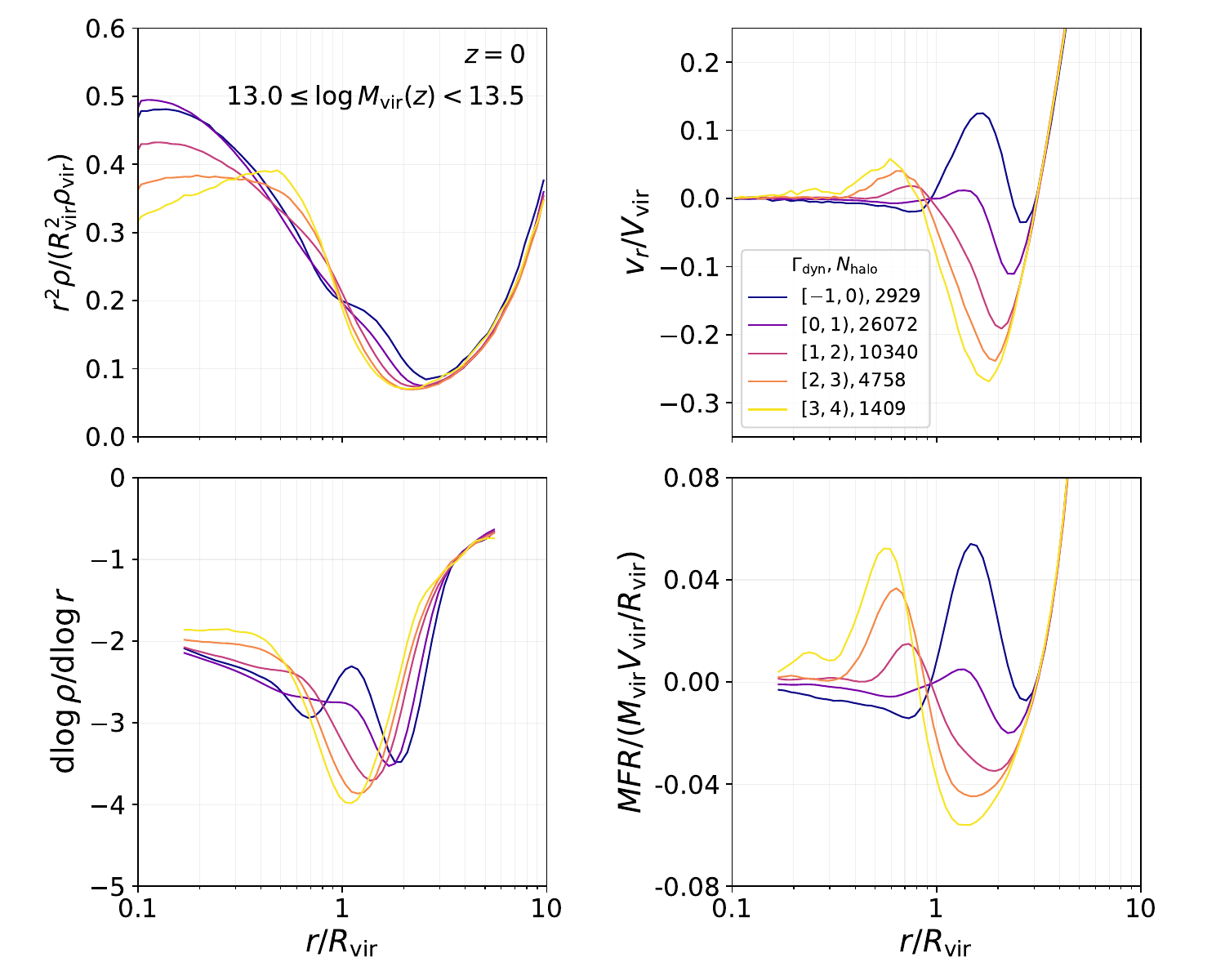}
\caption{Median radial profiles of dark matter halos in different bins of $\Gdyn$. The panels show the density profile (top left), logarithmic density slope (bottom left), radial velocity profile (top right), and mass flow rate (MFR; bottom right), with radial distance scaled by $\rvir$.  Halos are selected in the mass range $13 \le \log[\mvir/(\hM)] < 13.5$ at $z=0$ and are divided into bins of $\Gdyn$. The corresponding $\Gdyn$ ranges and number of halos in each bin are indicated in the legend. All quantities are normalized by the corresponding virial values.
\label{fig:prof_on_mar}}
\end{figure*}

Figure~\ref{fig:prof_on_mar} shows the normalized profiles of group-mass halos in different $\Gdyn$ bins. All profiles exhibit systematic variations with $\Gdyn$. The dependence of the density slope profile on $\Gamma$ has been investigated in detail in previous works \citep{DK14,More15}, where higher $\Gamma$ is associated with a steeper outer density slope and a smaller splashback radius. \rev{We should note that those works used $\rtwom$ as the rescaling radius, with $z=0.5$ as the initial epoch, while our dynamical-time definition corresponds to an initial redshift of $z\approx0.36$. Nevertheless, the qualitative trend remains the same.}

The MFR profile exhibits a closely related trend. Since $\mathrm{MFR}(r)$ combines the density and radial velocity fields, its overall shape largely traces coherent inflow and outflow motions. The dependence of the radial velocity profile on the mass accretion rate has been reported in \citet{DK14}. However, the extrema of the MFR and radial velocity profiles do not strictly coincide, reflecting the additional modulation by the local density distribution. This demonstrates that the depletion radius, defined by the minimum of the MFR profile, is connected to but not trivially determined by the velocity field alone. The qualitative similarity among the depletion radius, the splashback radius, and the radius of maximum infall velocity suggests that these characteristic scales arise from a common underlying dynamical process.

The MFR profile therefore provides a complementary diagnostic of halo dynamics by encoding both the kinematic and density structure of the halo outskirts. Halos with higher $\Gdyn$ exhibit a more pronounced inflow, producing a deeper minimum in the MFR profile and a more contracted depletion radius. At the same time, enhanced outflow within the virial radius becomes apparent for high $\Gdyn$ halos, consistent with the presence of recently accreted material that has passed pericenter and is approaching splashback. In contrast, low $\Gdyn$ halos show weaker net mass transport within $\rvir$ and are closer to dynamical equilibrium, reflecting a quieter recent accretion history.

Beyond the virial radius, low $\Gdyn$ halos exhibit relatively elevated density and net outflow, indicative of a more extended mass distribution dominated by backsplash material rather than first infall. This shifts the minimum of the MFR profile to larger radii, leading to a larger depletion radius. Together, these trends provide a physical interpretation for the strong dependence of the depletion radius on the recent mass accretion rate. This motivates a quantitative characterization of the depletion radius as a function of the accretion rate, which we present in the following sections.

\subsection{Dependence on halo mass}\label{subsec:mass_dependence}

\begin{figure*}[t!]
\centering
\includegraphics[width=2\columnwidth]{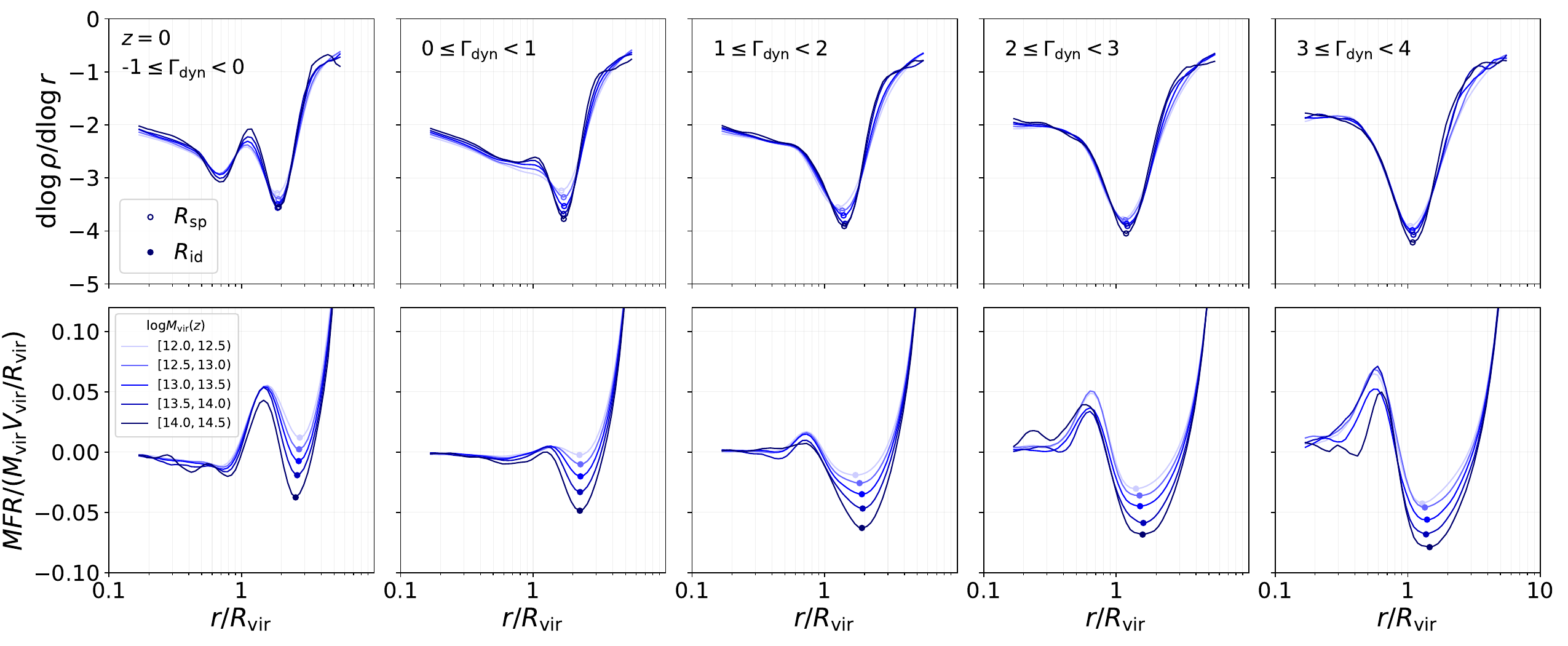}
\caption{Median density slope (top row) and MFR (bottom row) profiles at $z=0$ for halos in different $\mvir$ bins. Each column corresponds to a $\Gdyn$ bin, increasing from left to right. Different line shades denote different halo mass bins, as labeled in the legend. Unfilled and filled circles mark the splashback radius $\rsp$ and the depletion radius $\rid$, respectively, measured from the median profiles.
\label{fig:prof_on_m}}

\centering
\includegraphics[width=1.6\columnwidth]{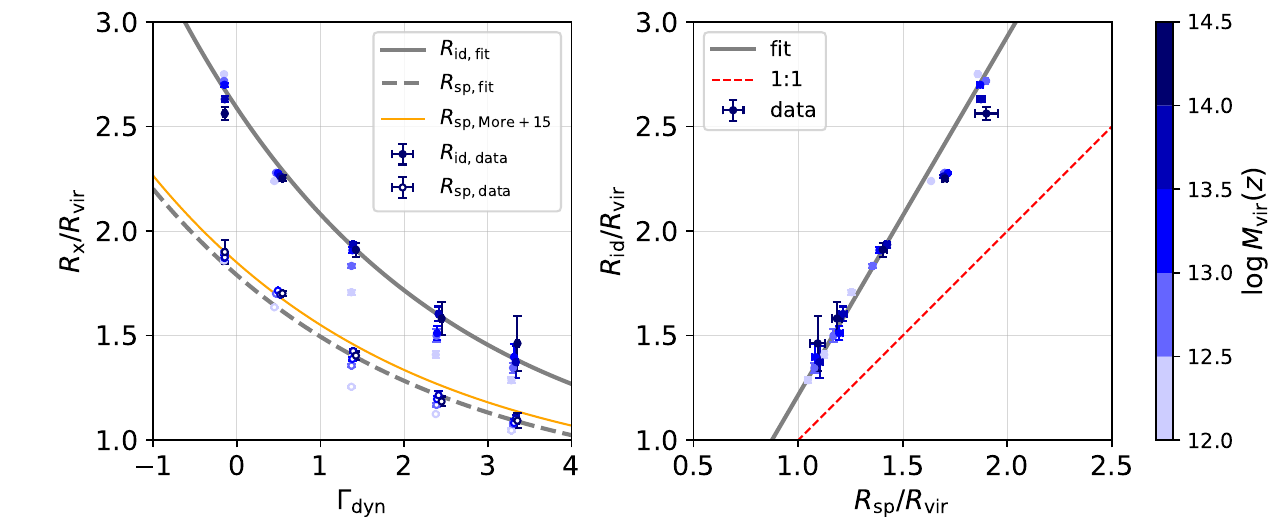}
\caption{\textit{Left}: $\rsp/\rvir$ (unfilled circles) and $\rid/\rvir$ (filled circles) as functions of $\Gdyn$ at $z=0$ for different $\mvir$ bins. Measurements are obtained from the median profiles in Figure~\ref{fig:prof_on_m}. gray curves show the best-fit relations derived in this work, and the orange curve shows the fitting formula of \citet{More15} converted to our virial radius definition. 
\textit{Right}: Relation between $\rid/\rvir$ and $\rsp/\rvir$ for the same sample. The gray line shows the fitted linear relation. \rev{The red dashed line shows the 1:1 relation for reference.}
\label{fig:r_on_m}}
\end{figure*}

Having established that the depletion radius strongly correlates with the mass accretion rate, we now examine whether this dependence varies across different halo masses. Figure~\ref{fig:prof_on_m} shows the median normalized profiles of halos binned by $\mvir$ within fixed $\Gdyn$ bins. Within each accretion rate bin, the profiles of different halo masses exhibit a high degree of self-similarity. The locations of both the splashback radius ($\rsp$) and the depletion radius ($\rid$), marked by unfilled and filled circles, respectively, remain nearly unchanged across the mass bins.

Although the locations of the characteristic radii are insensitive to halo mass, the amplitudes of the profiles show a mild mass dependence. In particular, the minimum of the normalized MFR profile becomes deeper for more massive halos at fixed $\Gdyn$, indicating a relatively stronger net mass transport. This is consistent with the overall expectation that more massive halos form later and are accreting more actively at the current epoch than lower mass ones.

A similar but weaker mass dependence is visible in the slope profiles, where the steepest slope becomes slightly more pronounced for more massive halos, especially at low $\Gdyn$. This is also consistent with the fact that the more massive halos have deeper MFR profiles. For these halos, more material has been added in the growth layer between the virialized region and the depletion radius, which marks the static point in the density evolution, creating a steeper outer profile.

Taken together, these results indicate that halo mass mainly modulates the amplitude and small scale structure of the outer profiles, while the locations of the characteristic boundaries relative to the virial radius are set by the recent mass accretion rate.

To quantify the dependence of the halo boundaries on accretion rate, we extract $\rid/\rvir$ and $\rsp/\rvir$ from the stacked profiles and present them as functions of $\Gdyn$ in the left panel of Figure~\ref{fig:r_on_m}. \rev{To estimate the statistical uncertainties of our measurements, we perform bootstrap resampling by randomly drawing with replacement within each bin, maintaining the same sample size as the full sample. For each bootstrap realization, we re-estimate the characteristic radii and the median $\Gdyn$ . This procedure is repeated 200 times. In Figure~\ref{fig:r_on_m} and all subsequent figures, the data points represent the results obtained from the full sample, while the error bars indicate the standard deviation derived from the 200 bootstrap realizations.}

As shown in Figure~\ref{fig:r_on_m}, both $\rsp/\rvir$ and $\rid/\rvir$ decline monotonically with increasing $\Gdyn$ and show only a weak dependence on halo mass. Following \citet{DK14} and \citet{More15}, we find that the overall dependence of the splashback radius on the accretion rate can be well described by an exponential form, 
\begin{equation}
\frac{R}{\rvir} = A + B \mathrm{e}^{-\Gdyn/C}. \label{eq:fitformula}
\end{equation}
As the mass dependence is negligible, we fit a common relation to the median radius of all the mass bins at each accretion rate, to find
\begin{equation}
\frac{\rsp}{\rvir} = 0.75 + 1.04 \mathrm{e}^{-\Gdyn/3}. \label{eq:rsprvir}
\end{equation}
\rev{The fitting is performed by minimizing the squared residuals between the model and the median radius across masses in each $\Gdyn$ bin, effectively assigning equal weight to all bins in order to capture the global shape of the relation. We have further verified that using inverse-variance weighting based on the bootstrap uncertainties yields very similar best-fitting parameters.}

This model is close to that obtained in \citet{More15} as shown in Figure~\ref{fig:r_on_m}, where we have corrected for the different definition of the virial radius in their original fitting (see Appendix~\ref{app:r200mrvir}). The small discrepancy is likely attributed to the different time intervals adopted to compute $\Gamma$: while \citet{More15} used $z=0.5$ as the initial epoch, our dynamical-time definition corresponds to an initial redshift of $z\approx0.36$.

As shown in the right panel of Figure~\ref{fig:r_on_m}, the depletion radius and the splashback radius are tightly correlated, following an approximately linear relation described by
\begin{equation}
\frac{\rid}{\rvir} = -0.50 + 1.72 \frac{\rsp}{\rvir}. \label{eq:ridrsp0}
\end{equation}
Combining this with Equation~\eqref{eq:rsprvir}, the depletion radius is then related to the accretion rate through
\begin{equation}
\frac{\rid}{\rvir} = 0.79 + 1.79 \mathrm{e}^{-\Gdyn/3}. \label{eq:ridrvir0}
\end{equation}
The very similar functional dependence on $\Gdyn$ and good linear relation suggest that $\rid$ and $\rsp$ are closely related and likely trace different but tightly connected aspects of the same underlying dynamical process.

In the self-similar spherical collapse model, both the splashback and depletion radii are expected to be associated with the first caustic \citep{Han26}, where the density profile exhibits a sharp feature and the mass flow transitions from single-stream infall to a regime influenced by orbiting material. In this idealized limit, the two radii coincide. In simulations, however, particles originating from the same shell develop a substantial radial spread due to deviations from spherical symmetry, radial orbit instability, and mergers, producing an extended splashback region rather than an infinitesimally thin shell. Different diagnostics of this region therefore yield slightly different characteristic radii. While $\rsp$ is determined by the steepest change in the density slope, $\rid$ depends on both density and radial velocity through the mass flow rate, naturally leading to a somewhat larger characteristic scale.

\rev{Although the explicit mass dependence is weak compared to the dominant $\Gdyn$ dependence and is omitted from our empirical fitting, it nonetheless exhibits a subtle and interesting pattern. In the slow accretion regime, more massive halos tend to have slightly smaller $\rid/\rvir$, whereas the opposite trend is observed in the fast accretion regime. The similar trend also holds for $\rsp/\rvir$. We note that some previous studies utilizing $\rtwom$ normalization \citep[e.g.,][]{Mansfield17, Diemer17, Diemer20} reported a persistent negative dependence of the splashback radius on halo mass at fixed $\Gamma$. A direct comparison with those conclusions remains challenging because of the different choices of normalizing radii and, more importantly, the significant methodological differences. Specifically, we use stacked median profiles while they analyze individual halo dynamics. Nevertheless, as we will demonstrate in Section~\ref{subsec:major_merger}, this residual mass dependence in our sample can be primarily explained by the varying fraction of major mergers across different mass bins, rather than an intrinsic mass scaling of the halo boundary.}

\subsection{Dependence on redshift}\label{subsec:redshift_dependence}

\begin{figure*}[t!]
\centering
\includegraphics[width=2\columnwidth]{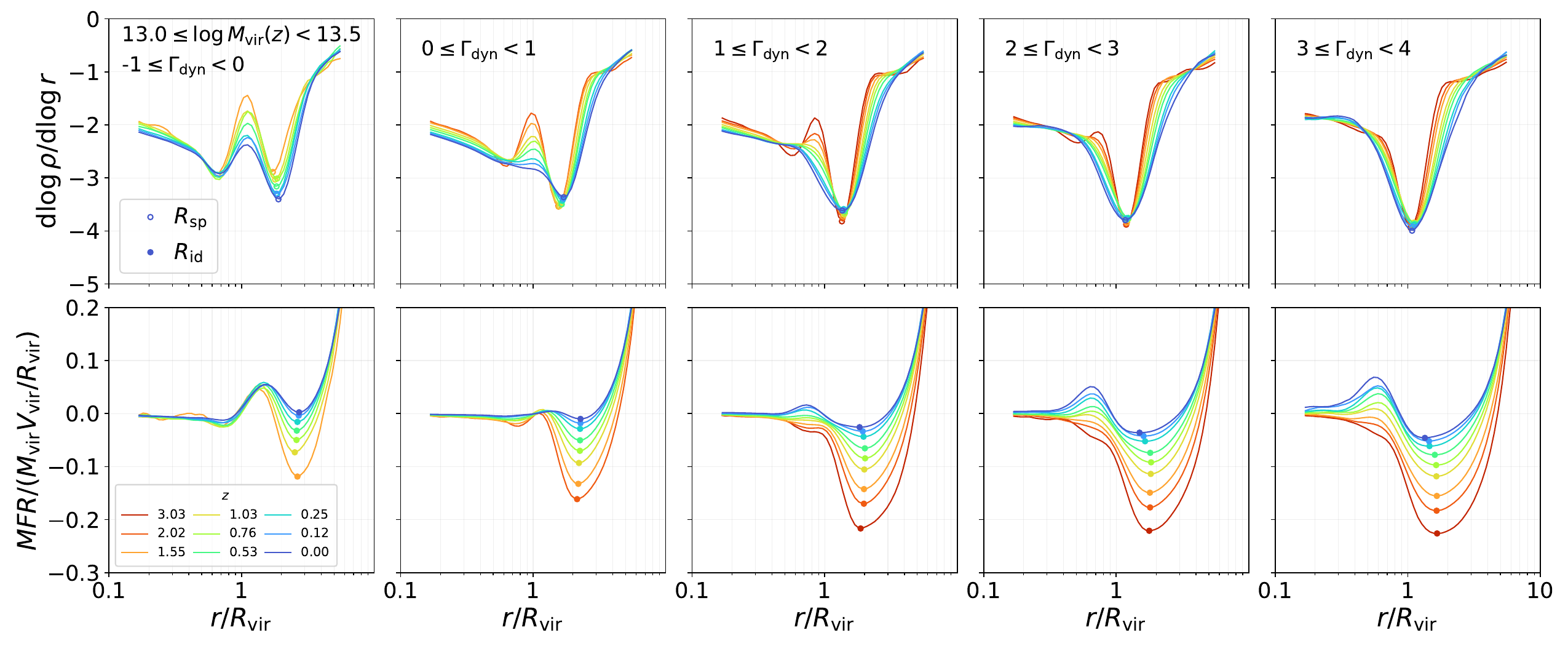}
\caption{Median density slope (top row) and MFR (bottom row) profiles of group-mass dark matter halos at different redshifts. Halos are selected at each redshift to have $13 \le \log[\mvir/(\hM)] < 13.5$ and are divided into bins of $\Gdyn$, shown in columns from left to right. Curves of different colors correspond to different redshifts, ranging from $z=3$ (red) to $z=0$ (blue). Radii are scaled by $\rvir$, and physical quantities are normalized by the corresponding virial values at each redshift. $\rsp$ (unfilled circles) and $\rid$ (filled circles) are measured from the median profiles.
\label{fig:prof_on_z}}

\centering
\includegraphics[width=1.6\columnwidth]{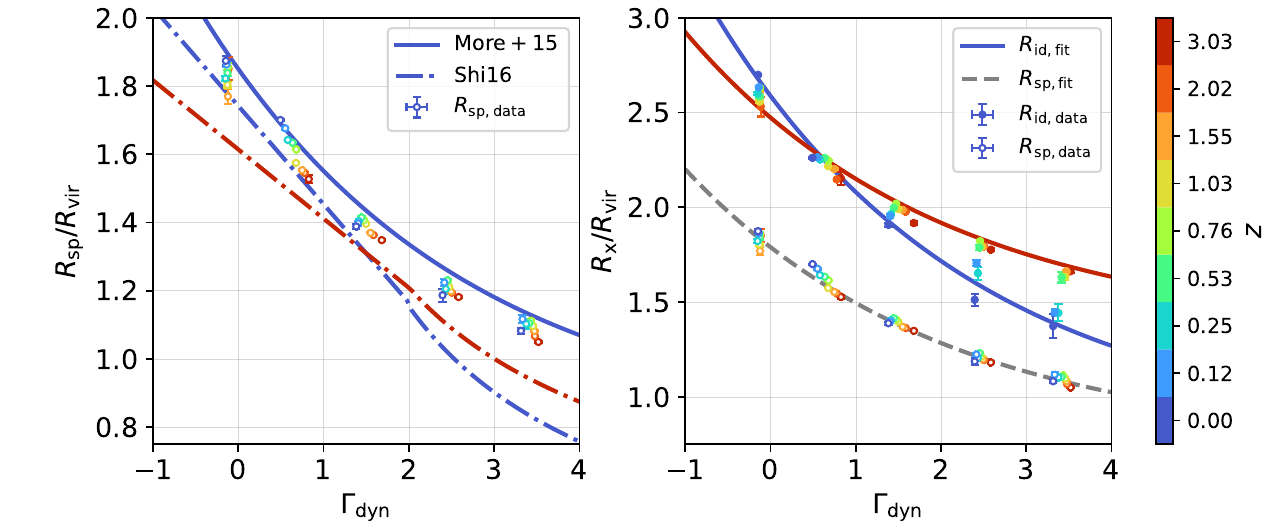}
\caption{Rescaled characteristic radii as functions of $\Gdyn$ at different redshifts. Colors indicate redshift. For each redshift and $\Gdyn$ bin, the plotted $\Gdyn$ and radii are obtained by taking the median of the measurements across all $\mvir$ bins. 
\textit{Left}: $\rsp/\rvir$ (unfilled circles). The solid curves show the fitting formula of \citet{More15}, originally defined for $\rsp/\rtwom$, converted to $\rsp/\rvir$ at two representative redshifts $z=0$ and $z=3$. The dash-dotted curves show the prediction of the self-similar spherical collapse model \citep{Shi16} at the same redshifts. 
\textit{Right}: $\rsp/\rvir$ (unfilled circles) and $\rid/\rvir$ (filled circles). Solid curves show the best-fit relations for $\rid$ at $z=0$ and $z=3$, while the dashed gray curve shows a single best-fit relation for $\rsp$ obtained by fitting all redshift data simultaneously.
\label{fig:r_on_z}}
\end{figure*}

In addition to the mass accretion rate, the self-similar spherical collapse model predicts an explicit redshift dependence of the splashback radius \citep[e.g.,][]{Adhikari14, Shi16}. In Figure~\ref{fig:prof_on_z}, we examine the evolution of the median normalized density slope and MFR profiles of group-mass halos, binned by $\Gdyn$ at each redshift. When radii are rescaled by $\rvir$ at the corresponding epoch, both profiles exhibit a high degree of self-similarity over the redshift range $0 \le z \le 3$, particularly for the density slope profiles, whose shapes and characteristic radii vary only mildly with redshift.

Despite this approximate self-similarity in the radial structure and in the locations of $\rsp$ and $\rid$, the MFR profiles show systematic redshift-dependent variations in amplitude. At higher redshift, the normalized MFR trough becomes progressively deeper and broader, indicating stronger net mass inflow over an extended radial range. This trend likely reflects the higher background density at earlier epochs, which enhances the inflow rate for halos with a given $\Gdyn$, even after normalization by $\rvir$ and the corresponding virial quantities.

More pronounced departures from self-similarity are found in the inner flow patterns. For example, in halos with $3 \le \Gdyn < 4$, the inner regions exhibit net outflow at low redshift, whereas at high redshift the same $\Gdyn$ bin is characterized by net inflow. This comparison suggests that a fixed value of $\Gdyn$ does not correspond to an identical dynamical state across cosmic time. A more detailed phase space analysis will be required to clarify the origin of this redshift dependence, which we leave for future work.

We extract $\rsp$ and $\rid$ from these profiles and present their $R/\rvir$--$\Gdyn$ relations at different redshifts in Figure~\ref{fig:r_on_z}. In the left panel, our measurements of $\rsp$ are compared with the empirical fitting model of \citet{More15}. We note that \citet{More15} adopted $\rtwom$ as the virial radius and reported a significant redshift dependence in the $\rsp/\rtwom$--$\Gamma$ relation. To enable a direct comparison, we convert their model to $\rsp/\rvir$ using the $\rtwom/\rvir$ ratio measured from our halo sample for each redshift (see Appendix~\ref{app:r200mrvir}). Interestingly, after this rescaling, the model exhibits only weak residual redshift dependence. This behavior is broadly consistent with our measurements, which show little systematic redshift evolution in $\rsp/\rvir$.

We further compare our results with the prediction of the self-similar spherical collapse model of \citet{Shi16}.\footnote{There is a typo in Equation (5) of \cite{Shi16}. The corrected version is (Xun Shi, private communication)
\begin{align}
    \frac{\rsp}{\rvir}= 
    \begin{cases}
    (-0.2 + 0.067 \ln\Om) s + 1.61 - 0.1\ln\Om, & s \leq 2,\\
    1.79\exp\left[(-0.66 + 0.2\Om)\ln s - 0.07 \Om\right], & s > 2,\nonumber
    \end{cases}
\end{align}
where $s$ is the theoretical mass accretion rate and $\Om$ is the matter density parameter at the corresponding redshift.}
The amplitude of the model is systematically lower than our measurements. This could be attributed to a systematic deviation in the theoretical accretion rate from the simulation measurements as discussed in \citet{Shi16}. The redshift evolution, however, is more pronounced in the model, in contrast to our measurements. Overall, the model predicts a flatter relation at a higher redshift.

In the right panel, we compare our $\rid$ measurements with $\rsp$. In contrast to $\rsp$, the depletion radius exhibits a clear and systematic redshift evolution. The trend of $\rid/\rvir$ with redshift qualitatively follows the expectation of \citet{Shi16}, even though $\rid$ is substantially larger than the predicted splashback radius. This suggests that $\rid$ may retain a stronger imprint of the cosmological background evolution than the splashback radius defined at the steepest density slope.

\begin{figure}[ht!]
\centering
\includegraphics[width=0.8\columnwidth]{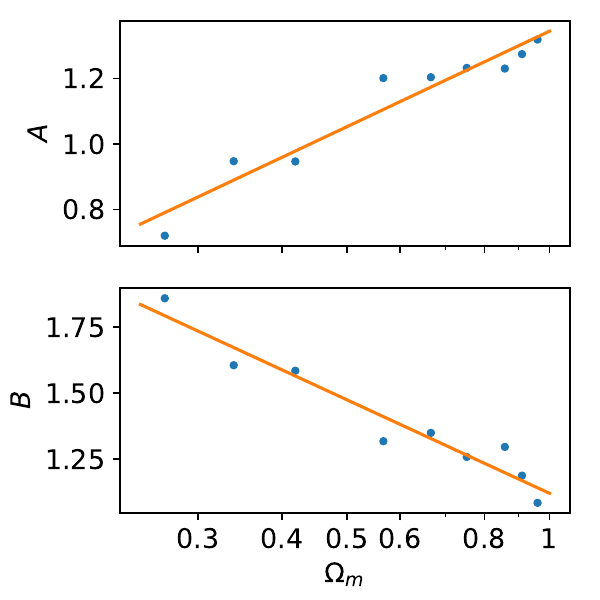}
\caption{Best-fit parameters $(A, B)$ of the $\rid/\rvir$--$\Gdyn$ relation (Equation~\eqref{eq:fitformula}) as functions of the matter density parameter $\Om$ at each redshift. The orange solid curves show linear fits in $\ln\Om$ to the data points.
\label{fig:ridfitparams}}
\end{figure}

To quantify this evolution, we fit the $\rid/\rvir$--$\Gdyn$ relation at each redshift using Equation~\eqref{eq:fitformula}. As in the $z=0$ case, we fix the parameter $C=3$ at all redshifts, while varying $C$ does not significantly improve the fit. Motivated by the spherical collapse framework, we parameterize the redshift dependence through the matter density parameter $\Om$. Figure~\ref{fig:ridfitparams} shows the best-fit parameters $(A, B)$ as functions of $\Om$. Both parameters are well described by linear relations in $\ln\Om$. We therefore obtain the following redshift-dependent parameterization:
\begin{equation}
\frac{\rid}{\rvir} = 1.35 + 0.42\ln\Om + (1.12 - 0.51\ln\Om) \mathrm{e}^{-\Gdyn/3}. \label{eq:ridrvir}
\end{equation}
Combining this expression with Equation~\eqref{eq:rsprvir}, we derive a redshift-dependent linear relation between $\rid$ and $\rsp$:
\begin{equation}
\frac{\rid}{\rvir} = 0.54 + 0.79\ln\Om + (1.07 - 0.49\ln\Om) \frac{\rsp}{\rvir}. \label{eq:ridrsp}
\end{equation}
Setting $\Om=0.268$ at $z=0$ recovers Equations~\eqref{eq:ridrvir0} and \eqref{eq:ridrsp0}. Note that even though Equations~\eqref{eq:ridrvir} and \eqref{eq:ridrsp} are fit using data up to $z=3$, they are expected to be largely applicable to higher redshift where the universe remains matter dominated with $\Om\approx 1$.

\section{Discussions}\label{sec:discussions}

\subsection{Impact of major merger}\label{subsec:major_merger}

\begin{figure*}[p]
\includegraphics[width=2\columnwidth]{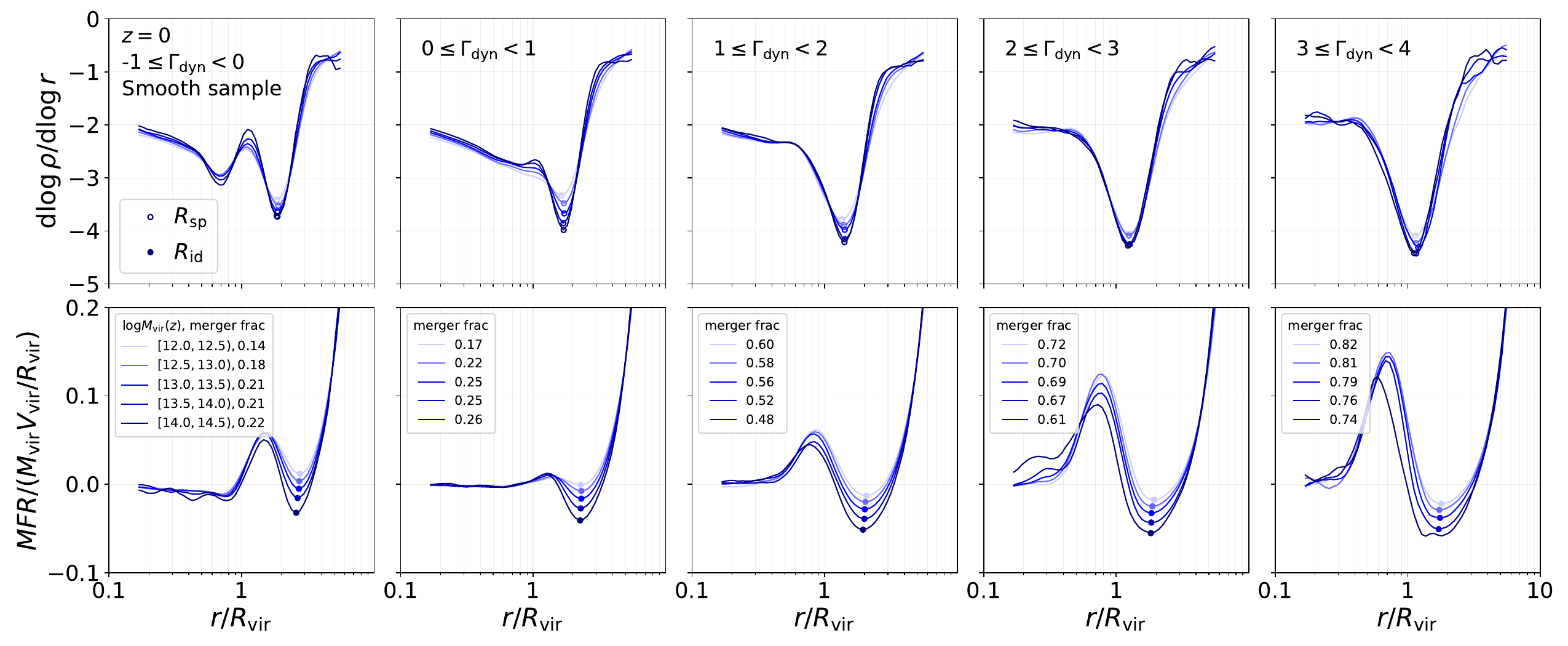}
\includegraphics[width=2\columnwidth]{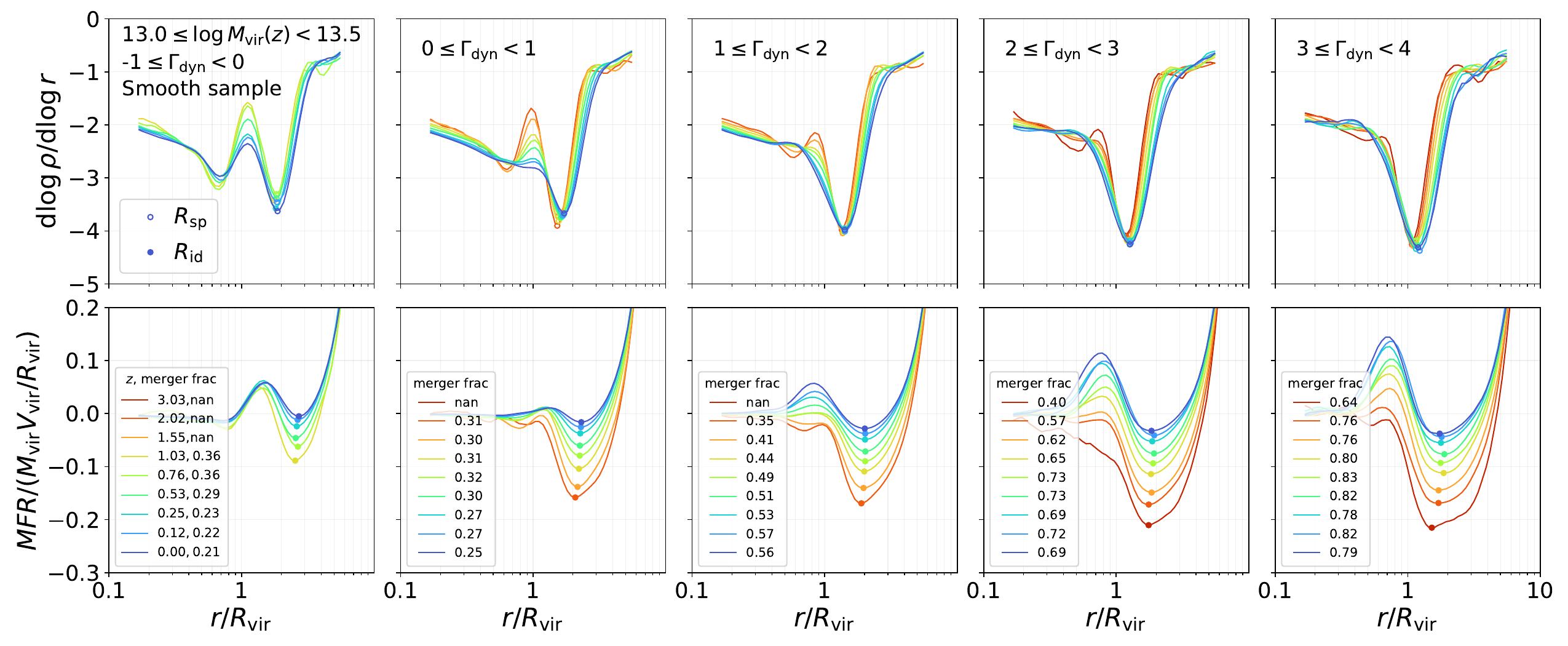}
\caption{Profiles of slope and MFR for halos excluding recent major mergers, analogous to Figures~\ref{fig:prof_on_m} and \ref{fig:prof_on_z}, \rev{with the major merger fraction within each bin labeled in the legend.} The slope profiles remain largely unchanged, whereas the MFR profiles show weaker net infall compared to the clean sample, especially for high $\Gdyn$ systems.
\label{fig:prof_dependence_smooth}}

\centering
\includegraphics[width=1.6\columnwidth]{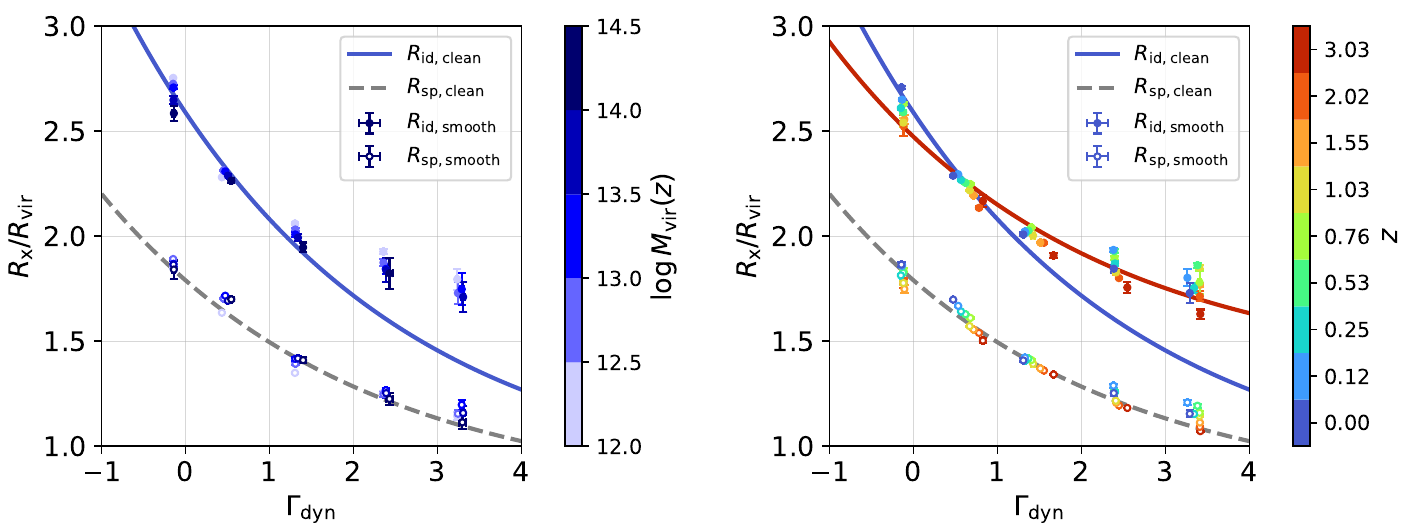}
\caption{Rescaled depletion and splashback radii as functions of $\Gdyn$ for the sample excluding halos with recent major mergers, corresponding to the left panel of Figure~\ref{fig:r_on_m} and right panel of Figure~\ref{fig:r_on_z}.\footnote{\rev{The highest-mass bin in the $3\le \Gdyn <4$ interval at $z=0$ exhibits a double-minimum feature, leading to a significant systematic uncertainty in the depletion radius measurement. It is therefore excluded from Figure~\ref{fig:r_dependence_smooth}.}} Lines show the fitting formula for the clean sample as in the right panel of Figure~\ref{fig:r_on_z}. Excluding major merger halos significantly increases $\rid$ in the fast accretion regime at low redshift and largely removes the redshift evolution of the $\rid/\rvir$--$\Gdyn$ relation.
\label{fig:r_dependence_smooth}}
\end{figure*}

Theoretical models of the halo boundary typically assume that the halo accretes smoothly with a given accretion rate. However, real halos grow not only via smooth accretion but also through mergers. In particular, major mergers can significantly perturb the mass and velocity distribution within halos, invalidating the smooth accretion as well as spherical symmetry assumptions. Moreover, the definitions of various spherical boundaries become problematic and can fluctuate significantly in the presence of major mergers. To assess the impact of major mergers on our results, we repeat the analysis excluding halos that experienced a recent major merger, defined as the accretion of a progenitor with a mass ratio greater than 1:9 within the past dynamical time, $t_{\mathrm{dyn}}$. Removing such systems allows us to isolate the dependence of halo boundaries on smoother accretion rather than transient merger-driven disturbances. \rev{In the following analysis, the sample obtained after excluding halos that experienced recent major mergers is referred to as the ``smooth sample''. The number of halos in each smooth sample bin is also listed in Table~\ref{tab:halo_number}. The fraction of removed major merger systems in each bin is labelled in Figure~\ref{fig:prof_dependence_smooth}.}

Figure~\ref{fig:prof_dependence_smooth} shows the density slope and MFR profiles for these smoothly accreting halos, while Figure~\ref{fig:r_dependence_smooth} presents the resulting radii. The slope profiles are only mildly affected, the depth becomes slightly deeper, consistent with \citet{DK14}. The splashback radius $\rsp$ increases slightly, particularly for high $\Gdyn$ halos.

In contrast, the MFR profiles show stronger responses. Excluding halos that have experienced recent major mergers reduces the depth of the inflow trough at all redshifts. Furthermore, for $\Gdyn \gtrsim 1$, the characteristic outflow associated with the splashback feature becomes more pronounced. This behavior indicates that, in the fast accretion regime, the accretion mode plays an important role in shaping the outer dynamical structure.

\rev{As shown in Figure~\ref{fig:r_dependence_smooth}, the redshift dependence of the $\rid/\rvir$--$\Gdyn$ relation is largely removed for the smooth sample. This is primarily due to an increase in $\rid/\rvir$ at the high $\Gdyn$ end at low redshifts, causing the low redshift relations to converge toward the high redshift ($\Om\approx1$) relation found for the clean sample. In contrast, the relation at the low $\Gdyn$ end remains nearly unchanged.}

\rev{The varying response of the halo boundary to the exclusion of major mergers across different accretion rate and redshift bins can be understood in terms of the differing major merger fractions within each bin. As shown in Figure~\ref{fig:prof_dependence_smooth}, at a given redshift, higher $\Gdyn$ values are strongly associated with higher major merger fractions. Excluding major merger systems therefore has a much stronger impact at the high $\Gdyn$ end than at the low $\Gdyn$ part. Similarly, within a given high $\Gdyn$ bin, halos at lower redshifts tend to have larger major merger fractions and are therefore more strongly affected by the removal of major merger systems. Note that the major merger fractions remain high in the high $\Gdyn$ regime even for the highest redshift bin. The relatively weak response to the exclusion of major mergers in these bins suggests that major mergers play a reduced role in determining $\rid$ at early epochs. The physical origin of this behavior remains unclear and merits further investigation in future work.}

\rev{The subtle mass dependence of $\rid/\rvir$ observed in Figure~\ref{fig:r_on_m} can also be understood in terms of the varying major merger fractions. As shown in the top two panels of Figure~\ref{fig:prof_dependence_smooth}, within the high $\Gdyn$ bins, low-mass halos exhibit systematically higher merger fractions than their more massive counterparts. Since low-mass halos generally have lower median accretion rates, reaching an exceptionally high $\Gdyn$ bin requires a more extreme recent growth history, which is more frequently driven by violent major mergers. Consequently, the higher merger fractions in low-mass halos suppress their stacked $\rid/\rvir$ at the high $\Gdyn$ end. This subtle mass dependence largely disappears once major mergers are removed.}

\rev{In contrast to the depletion radius, the splashback radius remains comparatively unchanged throughout these exclusions. This distinct behavior likely reflects the different responses of the halo profile on different scales to mergers. As the depletion radius is located further out into the non-equilibrium region around the halo, it is more sensitive to recent merger activities.}

Interestingly, the two radii for smoothly accreting halos can be approximately linked by a constant offset. Setting $\Om=1$ in Equation~\eqref{eq:ridrsp} yields 
\begin{equation}
    \rid \approx 0.54\rvir + \rsp, \label{eq:ridrspsmooth}
\end{equation}
which may serve as a useful starting point for unifying the two radii in theoretical models of halo collapse.

Despite the exclusion of major mergers, the lack of significant redshift evolution of both radii contrasts with the predictions of \citet{Shi16}, even though one might naively expect better agreement for these smoothly accreting halos. This discrepancy poses a challenge to our theoretical understanding of halo formation and calls for further investigation into the detailed impact of major mergers on halo structure, which we leave to future works.

\subsection{Dependence on other halo properties}\label{subsec:other_dependence}

To further assess whether the depletion radius is primarily governed by the mass accretion rate, we examine the dependence of $\rid/\rvir$ on several commonly used halo properties at $z=0$. In addition to $\Gdyn$, these properties include \citep[see][for detailed definitions]{Han19}:
\begin{itemize}
    \item the shape parameter, $e$, defined as the ratio between the largest eigenvalue and the sum of the three eigenvalues of the inertial tensor of the central subhalo;
    \item the spin parameter of the central subhalo, $j$, following the definition of ~\citet{1969ApJ...155..393P};
    \item the maximum circular velocity of the central subhalo, $V_{\mathrm{max}}$, normalized by the virial velocity, $\vvir$;
    \item the formation scale factor, $a_{1/2}$, at which the central subhalo acquires half of its current mass.
\end{itemize} Note all these properties are computed using particles in the central subhalo that corresponds to the smooth main component of the host halo, to avoid complications arising from substructures. The shape and spin parameters are expected to be shaped by the anisotropic accretion of the halo (e.g., through cosmic filaments) and the surrounding tidal field. The $V_{\mathrm{max}}/\vvir$ parameter is a common proxy for the halo concentration.

\begin{figure*}[ht!]
\centering
\includegraphics[width=2\columnwidth]{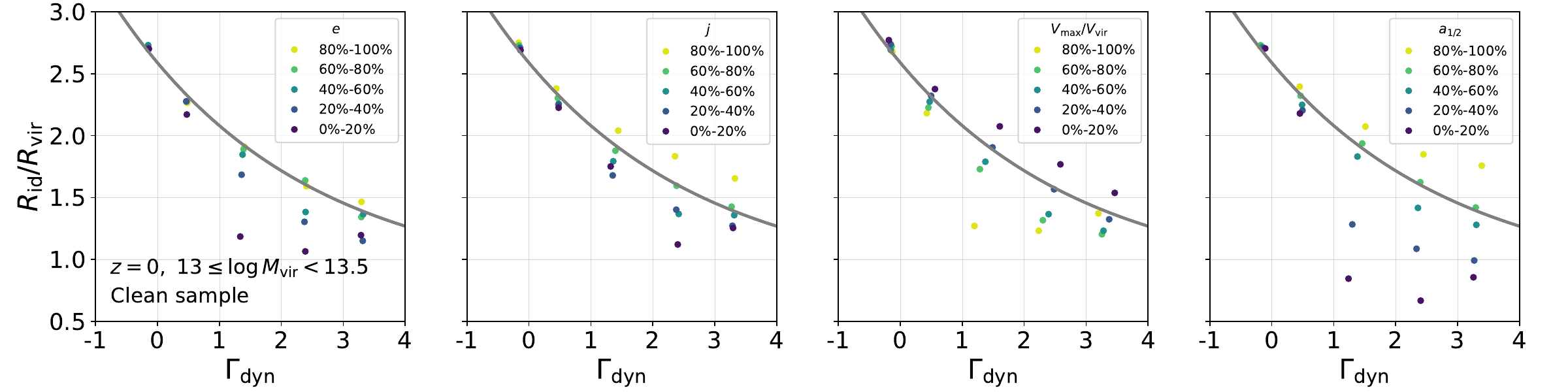}
\caption{Dependence of the $\rid/\rvir$--$\Gdyn$ relation on several secondary halo properties at $z=0$. Within each $\Gdyn$ bin, halos are further divided into five equal-number percentile bins (20\% each) according to increasing value of the following properties (from left to right): shape parameter $e$, spin parameter $j$, rescaled maximum circular velocity $V_{\mathrm{max}}/\vvir$, and half-mass formation time $a_{1/2}$. Colored points represent, for each percentile bin, the median $\Gdyn$ and $\rid/\rvir$ over all $\mvir$ bins. The solid gray curve shows the best-fit $\rid/\rvir$--$\Gdyn$ relation obtained without splitting the sample by secondary properties.
\label{fig:rid_on_prop}}

\centering
\includegraphics[width=2\columnwidth]{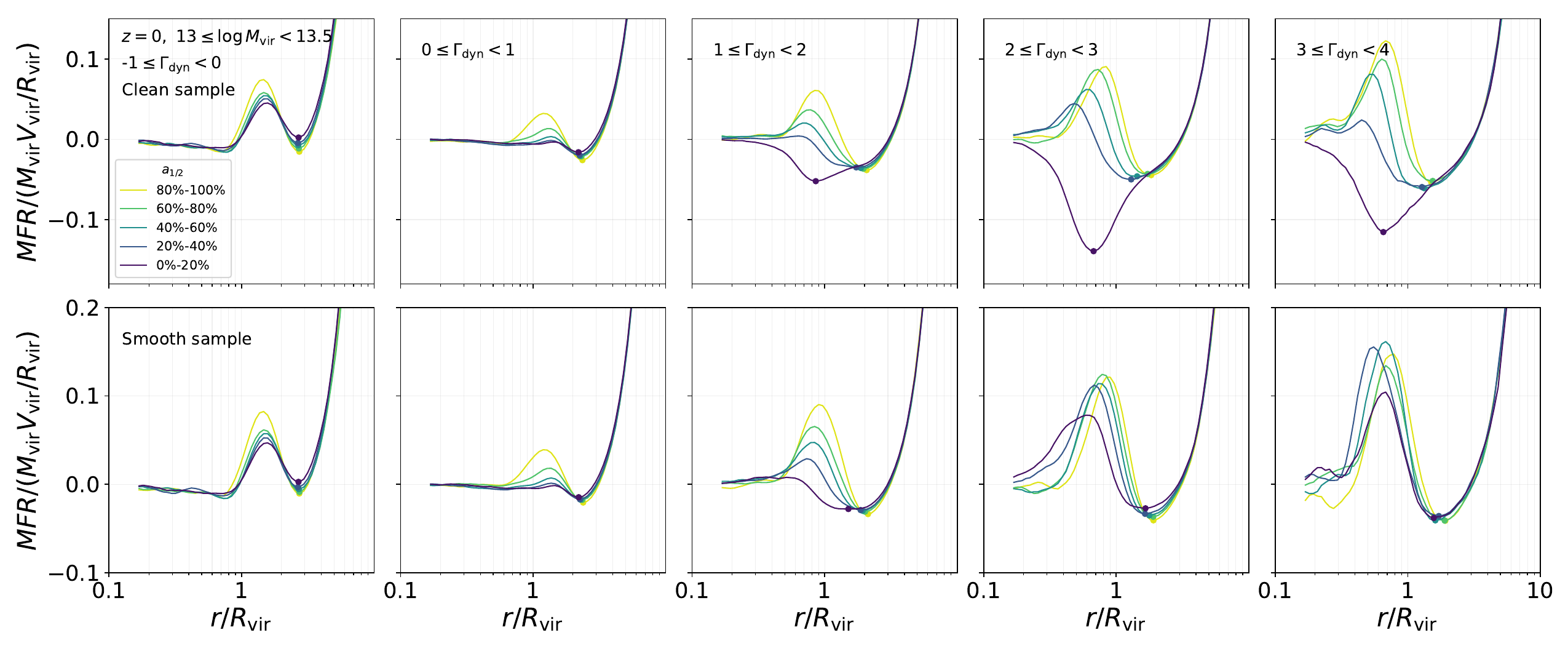}
\caption{Median MFR profiles at $z=0$ for halos with $13 \le \log[\mvir/(\hM)] < 13.5$, binned by $\Gdyn$ and by their half-mass formation time $a_{1/2}$. Halos in each $\Gdyn$ bin are divided into five equal-number percentile bins in increasing $a_{1/2}$. 
Top row: clean sample. Bottom row: smooth sample. 
Colored curves show the median MFR profile in each percentile bin, and colored points mark the corresponding $\rid$.
\label{fig:prof_on_ahalf}}
\end{figure*}

For each $\Gdyn$ bin, we further divide the halos into five equal-number percentile bins in each of the secondary properties above, and measure $\rid/\rvir$ from the stacked median profiles of each subsample. As shown in Figure~\ref{fig:rid_on_prop}, for $\Gdyn \lesssim 1$, the depletion radius shows little dependence on any parameter other than $\Gdyn$. For $\Gdyn \gtrsim 1$, however, additional dependences emerge. The variations associated with $e$, $j$, and $V_{\mathrm{max}}/\vvir$ are moderate, comparable to the redshift dependence discussed previously, while the dependence on $a_{1/2}$ is significantly stronger, with variations comparable to those induced directly by $\Gdyn$. Halos with smaller $a_{1/2}$ (earlier central formation) exhibit systematically smaller $\rid/\rvir$ at fixed $\Gdyn$. The dependence on $a_{1/2}$ is approximately a reversed version of the dependence on $V_{\mathrm{max}}/\vvir$, as the latter is a proxy of halo concentration and tightly anti-correlates with the formation time.

To understand the origin of the formation time dependence, we examine the stacked MFR profiles binned by $a_{1/2}$ for the clean sample in the top row of Figure~\ref{fig:prof_on_ahalf}. In the low $\Gdyn$ regime, the MFR profiles of different $a_{1/2}$ percentiles are similar, consistent with the weak secondary dependence in $\rid/\rvir$. In contrast, in the high $\Gdyn$ regime, halos with low $a_{1/2}$ show significantly stronger net inflow near the halo boundary and a suppressed or absent outflowing bump interior to the infall region. As a result, the depletion radius shifts inward for these systems.

This behavior can be naturally understood once we note that $a_{1/2}$ is defined using the bound mass of central subhalo, and therefore primarily traces the growth timescale of the relaxed inner component. At fixed high $\Gdyn$, a small $a_{1/2}$ implies that the central assembled most of its mass early and has grown relatively slowly thereafter. If the total halo mass is nevertheless increasing rapidly at the present time, this growth must predominantly occur through the addition of massive satellites rather than smooth accretion onto the central. In other words, the halo growth is dominated by (major) mergers. 

Such merger-driven growth enhances coherent inflow and disrupts the formation of a well-developed outgoing component, consistent with the MFR patterns observed for small $a_{1/2}$ halos in the high $\Gdyn$ regime. The resulting inflow dominated structure naturally shifts $\rid$ inward. This interpretation is fully consistent with the merger analysis presented in Section~\ref{subsec:major_merger}, where recently merged halos exhibit similar MFR features and reduced depletion radii.

To further test this picture, we repeat the $a_{1/2}$ binned MFR analysis for the smooth sample excluding recent major mergers (bottom row of Figure~\ref{fig:prof_on_ahalf}). In this case, the strong inflow enhancement previously seen for small $a_{1/2}$ halos at high $\Gdyn$ largely disappears, and the MFR profiles within a fixed $\Gdyn$ bin become much more similar across different $a_{1/2}$ percentiles. The secondary dependence of $\rid/\rvir$ on $a_{1/2}$ is correspondingly weakened. This confirms that the strong $a_{1/2}$ dependence in the fast accretion regime is primarily driven by recent merger-dominated growth.

Overall, these results reinforce the picture that the recent accretion rate is the primary driver of the depletion radius in the slow accretion regime, where halo growth is relatively smooth and closer to the assumptions of spherical collapse models. In the fast accretion regime, however, the accretion mode becomes important. Properties such as the formation time, concentration, shape, and spin of the central subhalo act as tracers of merger-driven or anisotropic growth, introducing additional structural variations beyond those captured by $\Gdyn$ alone. This reflects the increasingly complex dynamical state of rapidly growing halos. The depletion radius thus encodes not only the recent growth rate, but also the detailed modes of the growth.

\subsection{The universal factor $\rid \simeq 2\rvir$}\label{subsec:rid_2rvir}

\begin{figure*}[t!]
\centering
\includegraphics[width=2.2\columnwidth]{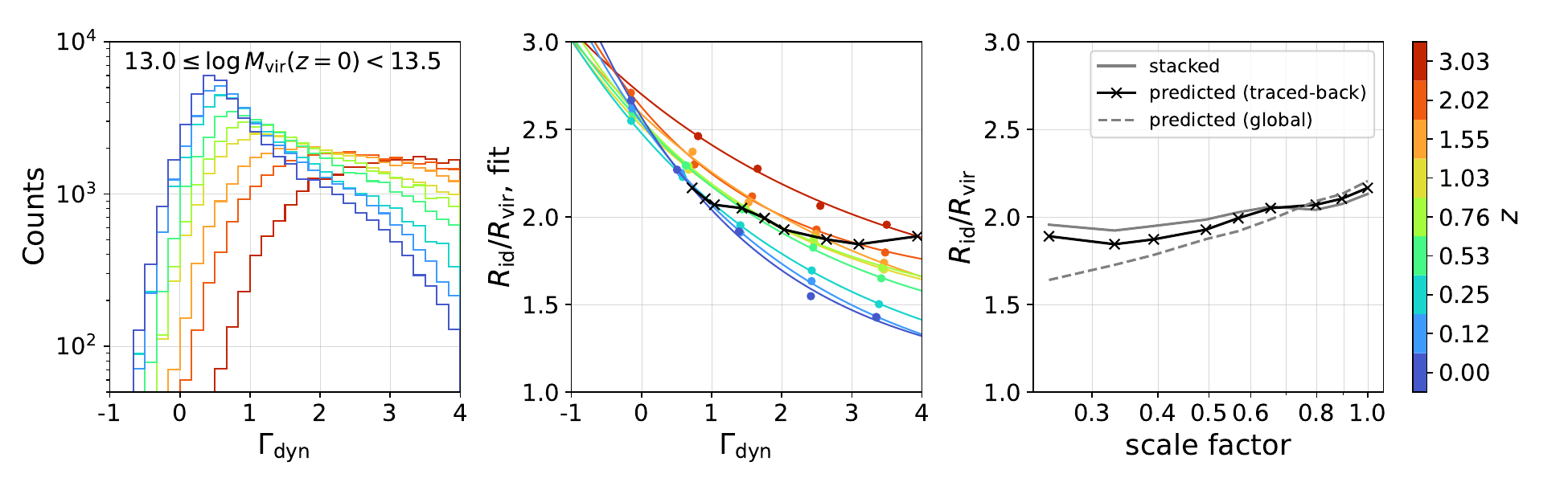}
\caption{Origin of the nearly constant ratio $\rid/\rvir \simeq 2$ across redshift. 
\textit{Left:} Distribution of $\Gdyn$ at each redshift for halos with $13 \le \log[\mvir/(\hM)] < 13.5$ at $z=0$, traced back along their main progenitor branches. 
\textit{Middle:} Predicted $\rid/\rvir$ at each redshift (crosses), obtained by mapping the median $\Gdyn$ at that redshift onto the corresponding best-fit $\rid/\rvir$--$\Gdyn$ relation (colored lines). Colored points show the measurements obtained from stacked MFR profiles at each redshift for the same traced-back halo sample shown in the left panel.
\textit{Right:} The predicted redshift evolution (crosses) generally matches the direct measurements from the stacked profiles of the traced-back sample (gray solid line). The gray dashed line shows the prediction by mapping the median $\Gdyn$ to the global $\rid/\rvir$--$\Gdyn$ relation (Equation~\eqref{eq:ridrvir}).
\label{fig:rid_2rvir}}
\end{figure*}

The depletion radius has been reported to maintain an approximately constant ratio $\rid/\rvir \simeq 2$ across a wide range of redshifts and halo masses \citep{FH21, Gao23}. However, this universality has to be a statistical behavior averaged over halos of different properties, given the dependence on accretion rate, redshift and other secondary properties of the depletion radius discovered above. We now verify this interpretation.

The left panel of Figure~\ref{fig:rid_2rvir} shows the distribution of $\Gdyn$ for halos traced back from a common $z=0$ mass bin with $13 \le \log[\mvir/(\hM)] < 13.5$. The median accretion rate increases systematically toward higher redshift. As shown previously, $\rid/\rvir$ decreases with increasing $\Gdyn$. However, at fixed $\Gdyn$ within the range populated by the majority of halos, $\rid/\rvir$ increases with redshift. When the median $\Gdyn$ at each redshift is mapped onto the corresponding $\rid/\rvir$--$\Gdyn$ relation (middle panel), these two trends partially compensate: higher redshift implies both higher accretion rates (which reduce $\rid/\rvir$) and an overall upward shift of the relation (which increases $\rid/\rvir$). The resulting prediction is a nearly constant $\rid/\rvir$ for the traced-back halo sample.

To compare with previous measurements \citep{Gao23}, we also compute $\rid$ directly from stacked MFR profiles at each redshift without binning by $\Gdyn$. As shown in the right panel of Figure~\ref{fig:rid_2rvir}, the $\rid/\rvir$ evolution predicted from the median accretion rate agrees well with that measured directly from the stacked profiles, with minor deviations at high redshift.

One complication in making the prediction lies in how the relation $\rid/\rvir$--$\Gdyn$ is constructed. Unlike Section~\ref{subsec:redshift_dependence}, here we derive the $\rid/\rvir$--$\Gdyn$ relation using only halos traced back from a common $z=0$ mass bin. For this traced-back sample, the depletion radius is systematically larger at high redshift than that shown in Figure~\ref{fig:r_on_z}, where the halo population is selected by mass independently at each redshift. Consequently, mapping the median $\Gdyn$ onto the global relation (Equation~\eqref{eq:ridrvir}) predicts a somewhat stronger redshift evolution (gray dashed curve in the right panel) than that obtained using the traced-back relation.

This subtle difference can again be interpreted as reflecting the different accretion modes of the traced-back and clean samples. At a fixed accretion rate at a high redshift, $z_1$, requiring the descendant to reach a fixed final mass at $z_0=0$ places a non-trivial constraint on the allowed growth histories between the two redshifts. As the infalling material outside $\rvir$ at $z_1$ affects the growth history afterwards, the above constraint implies a difference in the MFR profile between the traced-back and clean samples, resulting in their different depletion radii. This reinforces our previous finding that the depletion radius is sensitive to the detailed accretion mode of the halo beyond a single $\Gdyn$ parameter.

Nevertheless, the results above show that the empirical universality $\rid/\rvir \simeq 2$ does not arise from a fundamental physical constant. Instead, it emerges from the interplay between the evolution of the $\Gdyn$ dependence of the depletion radius and the systematic redshift evolution of the accretion rates.

\section{Conclusions}\label{sec:conclusions}
We investigated the dependence of the depletion radius ($\rid$) of dark matter halos on various halo properties using a large cosmological N-body simulation, and explored its physical origin through MFR profiles. Our main findings are summarized as follows:

\begin{enumerate} 
\item The depletion radius depends strongly on the recent mass accretion rate of the halo, $\Gdyn$, defined over the past dynamical timescale. Analogous to the splashback radius, halos with higher $\Gdyn$ tend to have smaller $\rid/\rvir$, indicating that rapid accretion leads to a more compact growth layer outside the virialized region.

\item At fixed $\Gdyn$, $\rid/\rvir$ shows only weak dependence on halo mass, but exhibits a mild yet systematic redshift evolution. We provide an accurate fitting formula (Equation~\eqref{eq:ridrvir}) describing $\rid/\rvir$ as a function of $\Gdyn$ and redshift. Compared with the splashback radius, the redshift dependence of $\rid$ is more pronounced when scaled by $\rvir$. The evolution trend is in better agreement with expectations from the spherical collapse model of \citet{Shi16} than that of the splashback radius: halos at higher redshift display a shallower $\rid/\rvir$--$\Gdyn$ relation and larger $\rid/\rvir$ at $\Gdyn\gtrsim1$. In contrast, although strong redshift evolution has been reported for $\rsp/\rtwom$ \citep[e.g.,][]{More15}, the evolution of $\rsp/\rvir$ is largely mitigated by the compensating redshift dependence of $\rtwom/\rvir$. Despite these differences, $\rid$ and $\rsp$ can be generally mapped to each other through a linear relation due to their shared dependence on the accretion rate.

\item The MFR profile provides direct physical insight into the origin and parameter dependence of $\rid$. The MFR pattern varies systematically with $\Gdyn$, halo mass, and redshift. Increasing $\Gdyn$ deepens and broadens the inflow region, shifts the radius of maximum inflow inward, and leads to a smaller $\rid/\rvir$. At fixed $\Gdyn$, the MFR profile depends only weakly on halo mass, although more massive halos exhibit slightly deeper inflow features even after normalization by virial quantities, indicating mild deviations from strict self-similarity. Higher redshift halos show broader and deeper inflow regions, whereas low redshift halos exhibit more prominent outflow features associated with splashback. These trends indicate that the mapping between $\Gdyn$ and the outer dynamical structure evolves with cosmic time.

\item Major mergers play an important role in shaping the accretion rate dependence of the depletion radius and are responsible for most of its redshift evolution. Merger-driven accretion perturbs the dynamical structure, enhances coherent inflow and suppresses the development of an outgoing component, leading to a systematically smaller depletion radius, especially in the fast accretion regime. When such systems are excluded, the redshift evolution of the $\rid/\rvir$--$\Gdyn$ relation is largely suppressed, and the relation approximately follows that at high redshift where $\Om\approx1$. In this case, $\rid$ and $\rsp$ differ by an approximately constant fraction of the virial radius (Equation~\eqref{eq:ridrspsmooth}).

\item The accretion rate is the dominant dependence of the depletion radius for slowly accreting halos with $\Gdyn\lesssim1$, where halo growth is typically smooth and close to the assumptions of spherical collapse. For rapidly accreting halos with $\Gdyn\gtrsim1$, however, $\rid$ exhibits additional dependence on halo properties including shape, spin, concentration, and formation time of the central subhalo. These secondary dependences suggest that the detailed accretion mode becomes important in the fast accretion regime. In particular, $a_{1/2}$ traces the relative growth of the relaxed central component. At a fixed growth rate of the total halo mass, $a_{1/2}$ thus acts as a diagnostic of the dominance of mergers in the mass growth, which in turn influences $\rid$. The additional dependences on shape and spin can be understood in a similar way, as they trace anisotropic or merger-driven accretion. These results indicate that $\rid$ encodes information about the detailed accretion mode beyond the recent accretion rate.

\item The approximately universal ratio $\rid/\rvir\simeq 2$ reported in previous works can be naturally explained by combining the $\rid/\rvir$--$\Gdyn$ relation with the redshift evolution of the typical accretion rate for halos traced back from a common $z=0$ mass bin. The opposing trends of the evolving median $\Gdyn$ and the redshift dependence of the $\rid/\rvir$--$\Gdyn$ relation largely compensate each other, leading to a nearly constant ratio over cosmic time. This compensation works particularly well for the traced-back sample, reflecting the distinct accretion modes of the traced-back and clean samples at high redshift.
\end{enumerate}

These findings help to establish the depletion radius as a useful diagnostic of the growth rate and dynamical state of the halo, providing a solid alternative to the splashback radius for observing the accretion rate. At the same time, our analysis also reveals additional sensitivity of the depletion radius to the detailed accretion mode and even the full accretion history of the halo in the fast accretion regime. It is interesting to note that the shock boundary in the gas component of the halo is also found to be sensitive to anisotropic accretion and major mergers~\citep[e.g.,][]{Congyao20,Congyao21,Sen26}. Some further theoretical and simulation efforts are required to investigate and understand these sophisticated dependences, which may eventually enable us to also use the depletion radius as a probe of these additional physical processes.

One major challenge in understanding the various scaling relations and their connection to theoretical models lies in the fact that the reference virial radius, either $\rvir$ or $\rtwom$, is itself affected by the detailed accretion mode and dynamical state of the halo. In the presence of recent major mergers, the virial radius can be temporarily boosted due to the newly added non-virialized mass from the merger. This overestimation can be clearly seen in the MFR profiles in the high $\Gdyn$ regime, naturally resulting in an apparent decrease in the $\rid/\rvir$. Note even without major mergers, the $\rvir$ does not always lie in the region of zero mass flow rate, in contradiction to the physical expectation of virialization, challenging its utilization as a reference radius in the first place. In a follow-up work, we will investigate these challenges in more detail.

\begin{acknowledgments}
We acknowledge helpful discussions with Xun Shi and Benedikt Diemer. This work is supported by NSFC (12595312), National Key R \& D Program of China (2023YFA1607800, 2023YFA1607801), China Manned Space Program (CMS-CSST-2025-A04) and Office of Science and Technology, Shanghai Municipal Government (grant Nos. 24DX1400100, ZJ2023-ZD-001).
The computation of this work is done on the GRAVITY supercomputer at the Department of Astronomy, Shanghai Jiao Tong University.
\end{acknowledgments}

\appendix
\section{Halo catalog and selection statistics}
Table~\ref{tab:halo_number} details the total number of halos initially identified by the FoF algorithm, the number of halos surviving our basic selection criteria, and the smooth accretion sample size after the exclusion of major mergers across different mass bins and snapshots.

Overall, the fraction of halos removed by our basic selection criteria typically ranges between 10\% and 30\%, depending on the halo mass and redshift. The fraction of systems further excluded due to recent major mergers ranges from roughly 30\% to 80\%, which is also explicitly displayed within each parameter bin in Figure~\ref{fig:prof_dependence_smooth}.

\begin{table*}
\centering
\caption{Number of halos in each mass and redshift bin. For each entry, we list the original halo count, followed by the numbers of halos in clean and smooth samples, respectively.}
\label{tab:halo_number}

\begin{tabular}{c|ccccc}
\hline
\multirow{2}{*}{$z$} &
\multicolumn{5}{c}{$\log_{10}(\mvir/\hM)$} \\
\cline{2-6}
& 12.0--12.5 & 12.5--13.0 & 13.0--13.5 & 13.5--14.0 & 14.0--14.5 \\
\hline
0.00 & 467,619/360,736/256,423 & 165,148/131,784/87,013 & 55,538/45,838/28,311 & 16,428/14,046/8366 & 3832/3416/1981 \\
0.12 & 472,379/355,040/236,703 & 164,528/128,174/77,816 & 54,015/43,781/24,535 & 15,464/13,066/6985 & 3382/3014/1520 \\
0.25 & 473,931/349,520/228,539 & 162,219/124,581/73,694 & 51,811/41,609/22,404 & 14,247/11,999/6213 & 2832/2539/1234 \\
0.53 & 467,910/337,718/194,102 & 153,863/116,376/58,820 & 45,412/36,109/16,107 & 11,137/9380/3933 & 1818/1622/636 \\
0.76 & 452,937/325,149/169,486 & 142,428/107,264/48,304 & 39,144/31,231/12,107 & 8366/7099/2538 & 1080/959/302 \\
1.03 & 424,511/307,285/157,177 & 126,103/95,848/42,542 & 31,255/25,235/9653 & 5616/4816/1701 & 505/460/143 \\
1.55 & 346,476/255,682/110,876 & 88,558/69,118/25,645 & 16,933/14,188/4471 & 1925/1711/500 & 84/76/16 \\
2.02 & 261,451/198,071/74,409 & 55,419/44,775/14,563 & 7847/6715/1804 & 571/505/132 & 5/5/1 \\
3.03 & 105,206/84,344/24,208 & 13,547/11,541/2825 & 807/727/146 & 13/12/3 & 0/0/0 \\
\hline
\end{tabular}
\end{table*}

\section{Fitting to the $\rtwom/\rvir$}\label{app:r200mrvir}

\begin{figure}[b]
\centering
\includegraphics[width=0.68\columnwidth]{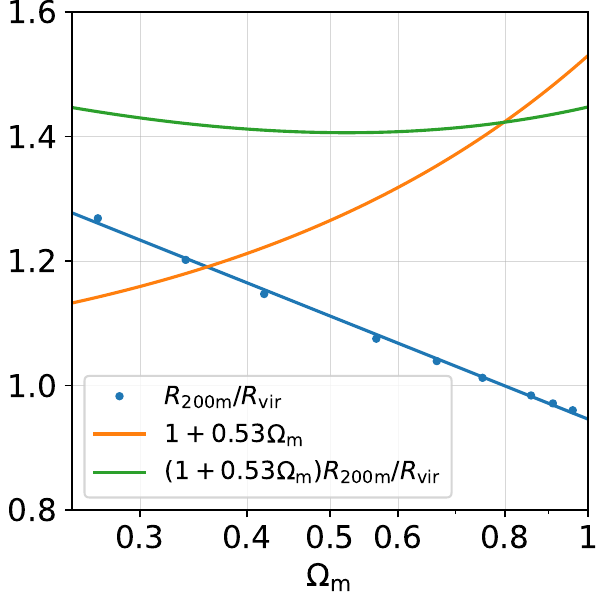}
\caption{Evolution of $\rtwom/\rvir$ (blue) and the $\Om$-dependent factor in the fitting function of \citet{More15} for $\rsp/\rtwom$ (orange). Blue points show the measured $\rtwom/\rvir$ at each redshift, and the blue line shows a linear fit to $\ln\Om$. The two terms evolve in opposite directions, and their product (green) remains nearly constant.
\label{fig:r200mrvir}}
\end{figure}

For each redshift, halo mass bin, and $\Gdyn$ bin, we measure the median value of $\rtwom/\rvir$. We find that at fixed redshift the variation across different mass and $\Gdyn$ bins is negligible compared with its redshift evolution. We therefore adopt, at each redshift, the median $\rtwom/\rvir$ over all bins, shown as the points in Figure~\ref{fig:r200mrvir}. This ratio follows a tight linear relation with $\ln\Om$, the best-fit result (blue line) is 
\begin{equation}
\frac{\rtwom}{\rvir} = 0.95 - 0.24\ln\Om .
\end{equation}

In the fitting formula of \citet{More15}, the redshift dependence of $\rsp/\rtwom$ enters explicitly through $\Om$. Converting to $\rsp/\rvir$ introduces an additional factor of $\rtwom/\rvir$. We find that the $\Om$-dependent term $1 + 0.53\ln\Om$ (orange) evolves in the opposite direction to $\rtwom/\rvir$, and their product (green) weakly depends on $\Om$. This cancellation explains why $\rsp/\rvir$ exhibits only weak redshift dependence in our analysis.

\bibliography{depletion_boundary}{}
\bibliographystyle{aasjournalv7}

\end{document}